\documentclass[prl,preprint,groupedaddress,amsmath]{revtex4} 
\usepackage{graphicx}
\usepackage{url}

\begin{document}

\title{Chemical-etch-assisted growth of epitaxial zinc oxide}

\author{E.~J. Adles} \affiliation{Department of Physics, NC State University, Raleigh, NC 27695-8202} \email[]{ejadles@ncsu.edu}
\author{D.~E. Aspnes} \affiliation{Department of Physics, NC State University, Raleigh, NC 27695-8202} 

\date{\today}
\begin{abstract}
We use real-time spectroscopic polarimetric observations of growth and a chemical model derived therefrom, to develop a method of growing dense, two-dimensional zinc oxide epitaxially on sapphire by metalorganic chemical vapor deposition. Particulate zinc oxide formed in the gas phase is used to advantage as the deposition source.  Our real-time data provide unequivocal evidence that: a seed layer is required; unwanted fractions of ZnO are deposited; but these fractions can be removed by cycling between brief periods of net deposition and etching.  The transition between deposition and etching occurs with zinc precursor concentrations that only differ by 13\%. These processes are understood by considering the chemistry involved.
\end{abstract}

\pacs{81.15.Gh,78.66.Hf} 
\keywords{zinc oxide, MOCVD, ellipsometry}
\maketitle

Zinc oxide (ZnO) is a transparent conducting oxide with a room-temperature band gap of 3.37 eV. It is currently under intense investigation for magneto-optic applications and as a cheap replacement for optical and optoelectronic devices currently depending on gallium and indium. \cite{Pearton2005aa,Klingshirn2007aa} Bulk, epitaxial, and nanostructured ZnO have been grown by a variety of methods including metalorganic vapor phase epitaxy, metalorganic chemical vapor deposition (MOCVD), molecular beam epitaxy, pulsed laser deposition, and vapor-liquid-solid processes. \cite{Pearton2005aa,Klingshirn2007aa,Look2001aa}  Of these processes MOCVD has several advantages, including the production of high-quality films through fine-tuning of various processing parameters,  simpler and less costly equipment, scalability, and higher throughput compared to conventional physical-vapor-deposition techniques. \cite{Malandrino2005aa}  

Here we  address two issues that cause difficulties in the growth of zinc oxide by MOCVD.  The common zinc precursors diethylzinc (DEZ) and dimethylzinc readily react in the gas phase with oxidizing species such as O$_2$, NO, and N$_2$O to create particles of sizes of the order of hundreds of nanometers or larger (the so-called particle problem). \cite{Hahn1998aa} Adverse effects of particulate ZnO in the gas phase range from deposition of poor-quality material to the clogging of process lines leading to reactor downtime.  Second, deposition of ZnO often yields granular or columnar structures and surface roughness of a few to a few tens of nm. \cite{Klingshirn2007aa}  Some of the recent attempts to address these issues include the use of exotic precursors such as Zn(TTA)$_2$TMEDA and Zn(TMHD)$_2$;\cite{Malandrino2005aa,Saraf2007aa}  alcohols as oxidizing agents;\cite{Hahn1998aa} modified reactor designs that separate precursors and effectively transform the growth process into alternating layer epitaxy;\cite{Du2005aa} atmospheric pressure deposition;\cite{Dai2006aa,Matthews2006aa} and variation of the VI/II ratio. \cite{Gruber2002aa}

In this work we report results obtained by real-time spectroscopic polarimetry that provide insights into the chemistry and growth of ZnO by MOCVD.  We take advantage of these insights to develop an efficient process where the consequences of the above difficulties are minimized.    Minimum modification of our MOCVD reactor is required.  We show that particulate ZnO in the gas phase, when properly managed, provides a useful source for deposition.

Our deposition system is a modified Emcore GS-3300 MOCVD reactor with an integrated real-time spectroscopic polarimeter for in-situ analysis and control of growth.  The polarimeter measures the relative reflectance and the complex reflectance ratio from 240 to 840 nm at a 4 Hz rate, allowing growth to be followed on this time scale.  Details of the growth system and optical diagnostics are discussed in previous work. \cite{Flock2003aa,Flock2004aa}  The only modification specific to ZnO growth was the addition of a down-tube to keep the precursors separated as long as possible.  The down-tube terminated approximately 2.5 cm above the sample, a spacing determined by the need to maintain an unobstructed light path for the polarimeter.

A qualitative understanding of the connection between ZnO density and the  $<\mspace{-5.0mu}\epsilon\mspace{-5.0mu}>$ spectra measured by polarimetry can be achieved by means of the linear expansion of the Fresnel equations for small $d / \lambda$, expressed in pseudo-dielectric function form \cite{Aspnes1976aa}:
\begin{equation}
	\begin{split}
	 <\mspace{-5.0mu}\epsilon\mspace{-5.0mu}> &= <\mspace{-5.0mu}\epsilon_1\mspace{-5.0mu}> + i <\mspace{-5.0mu}\epsilon_2\mspace{-5.0mu}> \\
	&= \frac{4 \pi i d}{\lambda}\frac{\epsilon_s (\epsilon_s - \epsilon_o) (\epsilon_o - \epsilon_a)}{ \epsilon_o (\epsilon_s - \epsilon_a)} \sqrt{ \frac{\epsilon_s}{\epsilon_a} - sin^2 \phi},
\end{split}
	\label{eq:linear-expansion}
\end{equation}
where $\epsilon_s$, $\epsilon_o$,  and $\epsilon_a = 1$ are the dielectric functions of the sapphire substrate, the ZnO overlayer, and the ambient, respectively, and $\phi = 70^\circ$ is the angle of incidence.  In the region of transparency of ZnO we have $\epsilon_s \approx 3.12$ and $\epsilon_o \approx 4.00$, neglecting birefringence and dispersion.  Thus if bulk ZnO is being deposited, $\epsilon_o > \epsilon_s$ and Eq. \eqref{eq:linear-expansion} shows that in the transparent region $<\mspace{-5.0mu}\epsilon_2\mspace{-5.0mu}>$ will be negative.  Conversely, if $<\mspace{-5.0mu}\epsilon_2\mspace{-5.0mu}>$ is positive, then $\epsilon_o < \epsilon_s$, i.e., the deposited ZnO contains a significant fraction of voids.

We report the results of two experiments where ZnO was deposited on (0001) sapphire (Al$_2$O$_3$) substrates.  In run \#1, we illustrate various aspects of growth using DEZ and O$_2$ as the zinc precursor and oxidizer, respectively. Ultrahigh purity nitrogen (UHP N$_2$) was used as the carrier gas for DEZ.  The flow rate of the carrier gas was manually adjusted between 10 to 100 sccm in response to real-time values of $<\mspace{-5.0mu}\epsilon_2\mspace{-5.0mu}>$. All other process settings remained constant. N$_2$ boiled off from liquid N$_2$ was used as a pump ballast, to maintain chamber pressure, and to prevent deposition on the optical viewports.  The growth parameters were: substrate temperature 391$^{\circ}$C; chamber pressure 80 Torr; O$_2$ flow rate 10 sccm;  main N$_2$ flow rate 1.5 slm; DEZ bubbler temperature -10$^{\circ}$C, and UHP N$_2$ (DEZ carrier gas) 10-100 sccm.  No pressure controller was used in the DEZ lines, so the pressure in the DEZ bubbler was established by the chamber pressure.

In run \#2 we apply the information obtained to achieve two-dimensional growth of dense material.  In addition to DEZ and O$_2$, 3\% NO in N$_2$ was added to attempt p-type doping by nitrogen substitution.  Parameters here were: substrate temperature 490$^{\circ}$C; chamber pressure 140 Torr; O$_2$ flow rate 20 sccm; 3\% NO in N$_2$ flow rate 100 sccm; main N$_2$ flow rate 1.5 slm; DEZ bubbler temperature 15$^{\circ}$C; and UHP N$_2$ (DEZ carrier gas) flow rate 50 sccm. In contrast to run \#1, we used a pressure controller in the DEZ line after the bubbler to set the pressure in the bubbler.  This allowed us to cycle between short periods of deposition and removal by varying the pressure in the bubbler.  This proved to be the key that allowed us to grow dense material.  Specifically, we cycled the pressure between 450 and 400 Torr, corresponding to DEZ flow rates of 1 and 1.125 sccm, respectively.  The cycle period was approximately 30 seconds, limited by the response time of the pressure controller.

\begin{figure}
\includegraphics[scale=0.6]{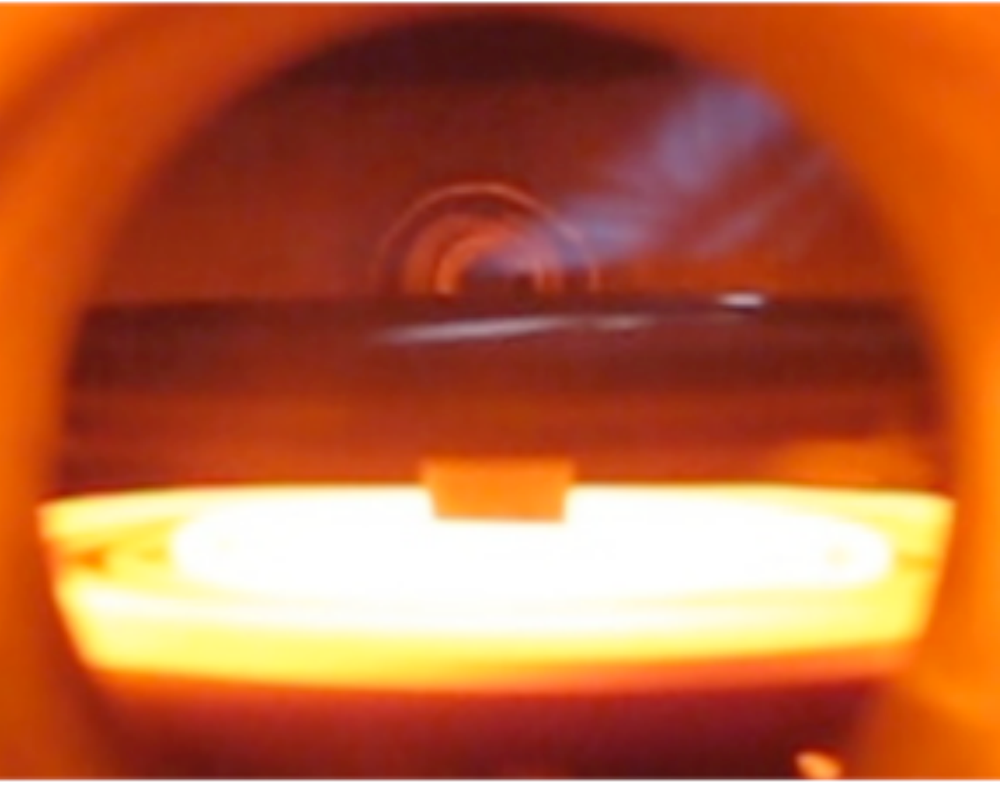}
\caption{View inside the growth  chamber showing a typical vertical distribution of ZnO particles.  The heater, substrate, and down-tube are visible at the bottom, center and top, respectively. The polarimeter beam entering at the upper right is made visible by scattering from the particles, indicating sizes of the order of hundreds of nm. (color online) \label{fig:virtual-source}}
\end{figure}

Figure \ref{fig:virtual-source} shows a typical vertical distribution of ZnO particles above the sample surface that results from the reaction of DEZ and O$_2$ in the gas phase.  Particles are made visible by light scattered from the incoming polarimeter beam.  The scattering indicates ZnO particle sizes of the order of hundreds of nm.  Of particular interest is the sharp cutoff of scattered light approximately 5 mm above the sample.  Within this 5 mm  the particle size has obviously dropped below the scattering threshold, indicating sublimation or etching.  Since the same process is likely to be occurring at the substrate, we conclude that fairly large exchange currents must be present, although with the net result that ZnO is deposited on the substrate.  Thus two exchange currents must be managed: that originating with the particles, and the second from the material deposited on the substrate.  In our system the susceptor provides the necessary heat to drive the reactions.  More generally a separate heating element could be used for independent control.

\begin{figure}
\includegraphics[scale=0.43]{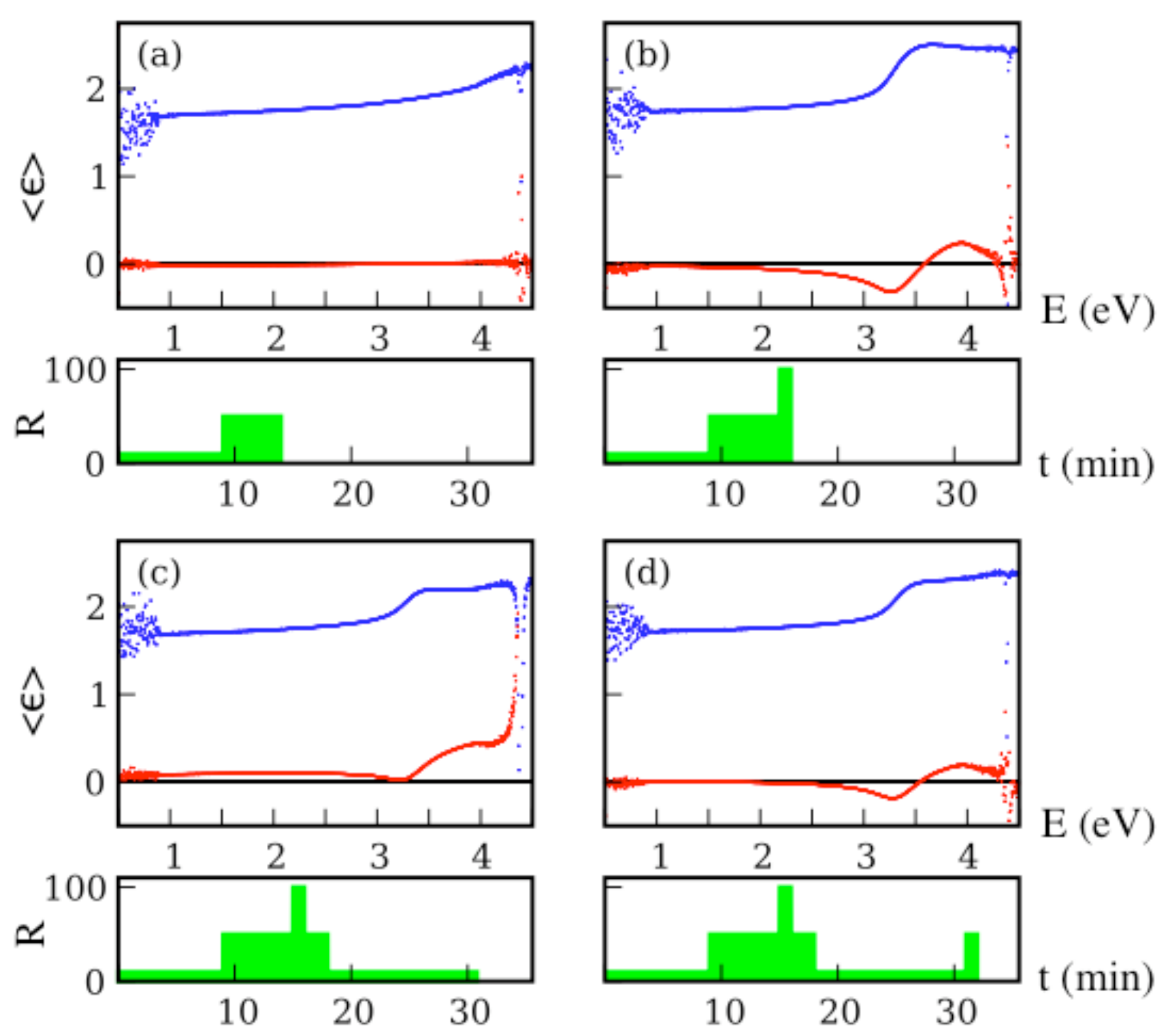}
\caption{Four subplots from run \#1 showing $<\mspace{-5.0mu}\epsilon\mspace{-5.0mu}>$ vs. energy in the top pane and the history of the DEZ carrier gas flow rate R in the bottom pane.  ZnO deposition is indicated by structure near 3.3 eV.  (a) shows that no ZnO deposition occurs at initial carrier flow rates of 10 and 50 sccm.  (b) ZnO deposition begins at 100 sccm.  (c) At 10 sccm the structure near 3.3 eV is reduced and $<\mspace{-5.0mu}\epsilon_2\mspace{-5.0mu}>$ below 3 eV increased, indicating sublimation of deposited ZnO and resulting porous material. (d) At 50 sccm the structure near 3.3 eV increases and $<\mspace{-5.0mu}\epsilon_2\mspace{-5.0mu}>$ below 3 eV becomes negative, indicating deposition and infilling of the porous material.  Note that the seed layer established in (b) at 100 sccm allows deposition at 50 sccm in (d) whereas it did not occur earlier in (a). \label{fig:first-run}}
\end{figure}

With the basic conditions established, we now consider ZnO deposition in detail. Figures \ref{fig:first-run} show $<\mspace{-5.0mu}\epsilon_1\mspace{-5.0mu}>$ and $<\mspace{-5.0mu}\epsilon_2\mspace{-5.0mu}>$ as a function of energy in eV, at four different times into run \#1.  These spectra correspond approximately to reflection and absorption, respectively. In Fig.~\ref{fig:first-run}(a) $<\mspace{-5.0mu}\epsilon\mspace{-5.0mu}>$  describes the dielectric response of the bare Al$_2$O$_3$ substrate, which is transparent and highly polished.  Therefore, no absorption is seen and $<\mspace{-5.0mu}\epsilon_1\mspace{-5.0mu}>$ is nearly flat across the visible spectrum.  The presence of any  ZnO would result in structure near 3.3 eV as seen in Figs.~\ref{fig:first-run}(b)-(d). From the lower pane of Fig.~\ref{fig:first-run}(a) we see that carrier gas flow rates of 10 and 50 sccm yield no initial deposition. At 100 sccm deposition begins, as indicated by the structure developed near 3.3 eV in Fig.~\ref{fig:first-run}(b). Here, $<\mspace{-5.0mu}\epsilon_2\mspace{-5.0mu}>$ is negative below 3.3 eV.  From Eq.~\eqref{eq:linear-expansion} we conclude that the deposited material has a refractive index n greater than that of Al$_2$O$_3$, which is characteristic of bulk ZnO.  Figure \ref{fig:first-run}(c) shows that reducing the flow rate to 10 sccm results in removal of the deposited ZnO.  This is evident by the reduction in structure near 3.3 eV, and that  $<\mspace{-5.0mu}\epsilon_2\mspace{-5.0mu}>$  below 3 eV has become positive.  This indicates that n of the overlayer is now less than that of Al$_2$O$_3$, ie., that the ZnO has become porous.  We interpret this as preferential sublimation or etching of either defective material or of crystal orientations other than that epitaxially grown on the substrate.  Figure \ref{fig:first-run}(d) shows that returning to 50 sccm  results in deposition, whereas no deposition initially occurred at 50 sccm as seen in Fig.~\ref{fig:first-run}(a).  In addition, the $<\mspace{-5.0mu}\epsilon_2\mspace{-5.0mu}>$ spectrum shows that this new material is filling in the voids. From this we deduce that a seed or striking layer must be established for subsequent growth to occur, and that a cyclic pattern of removal and deposition may result in improved material.  

\begin{figure}
\includegraphics[scale=0.5]{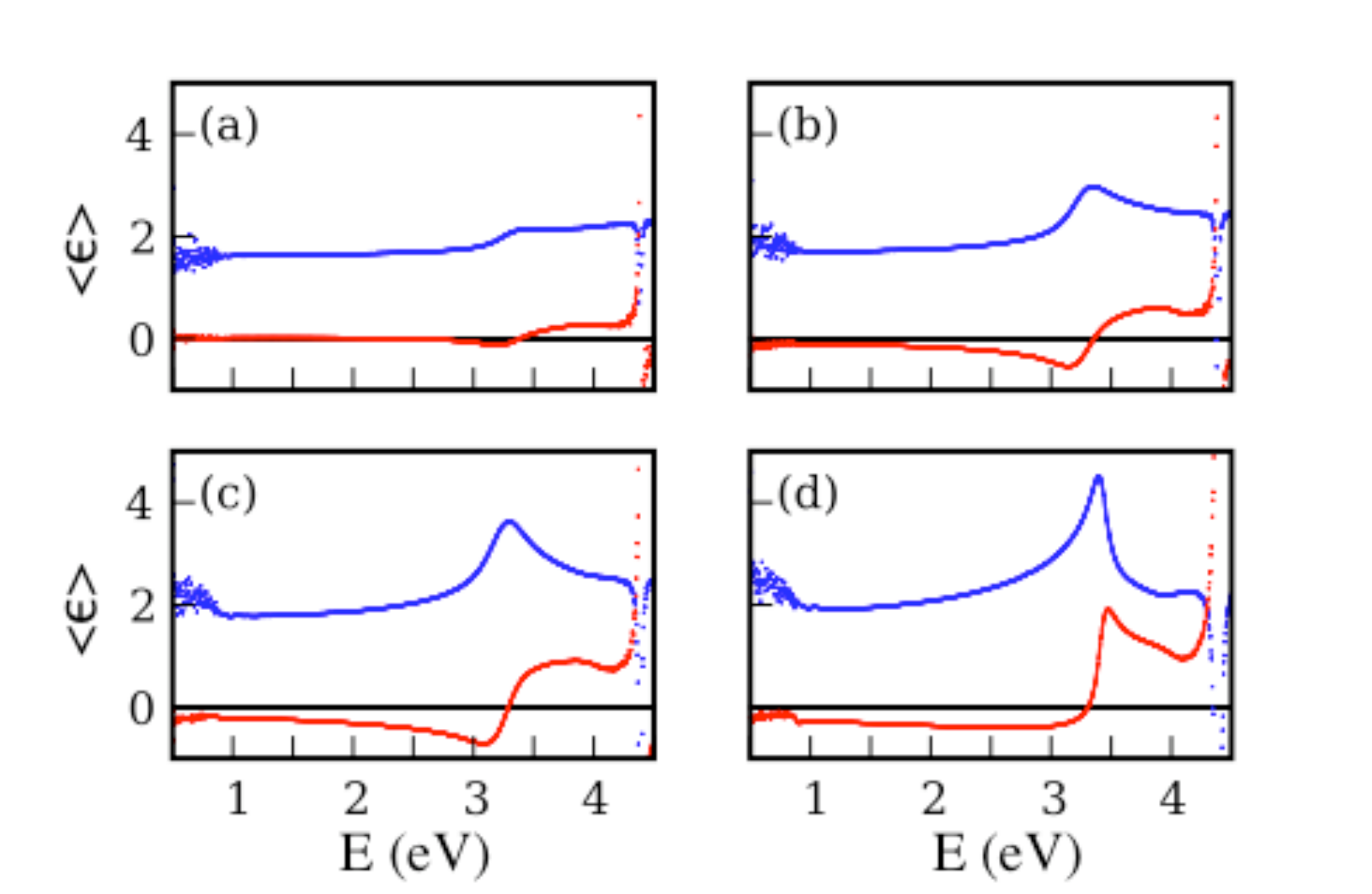}
\caption{Subplots (a)-(d) showing the progression with time of $<\mspace{-5.0mu}\epsilon\mspace{-5.0mu}>$ during run \#2, where the DEZ flow was cycled between 1.125 and 1 sccm for approximately 15 seconds each.   (a)-(c) Regular cycling between brief periods of net deposition and net sublimation results in increasingly well-defined structure near 3.3 eV and more negative values of $<\mspace{-5.0mu}\epsilon_2\mspace{-5.0mu}>$, both of which indicate high-quality material.  (d) illustrates the well-defined structure observed after cool down.
\label{fig:second-run}}
\end{figure}

\begin{figure}
\includegraphics{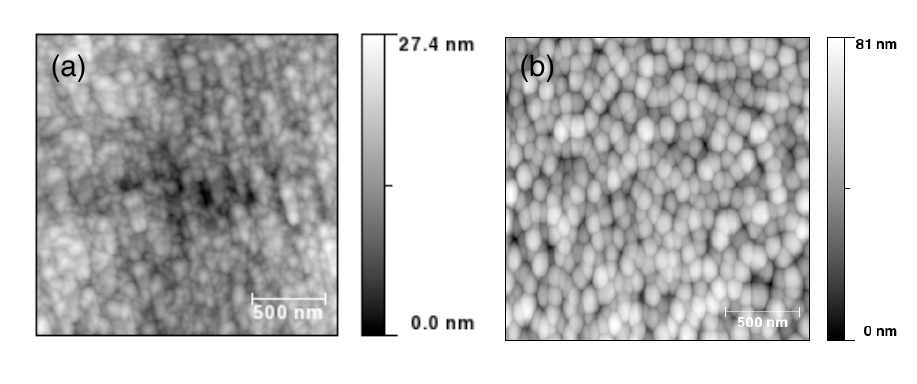}
\caption{AFM data for a ZnO film prepared with (a) and without (b) the cyclic growth process.  RMS roughness over a 2x2 $\mu$m$^2$ area is (a) 3.93 nm and (b) 12.36 nm.
\label{fig:afm}}
\end{figure}

Figures \ref{fig:second-run}(a)-(d) show the progression of  $<\mspace{-5.0mu}\epsilon\mspace{-5.0mu}>$ during run \#2, where the cyclic growth strategy suggested by the above results was implemented.  In Figs.~\ref{fig:second-run}(a)-(c) the structure near 3.3 eV is increasingly better defined and $<\mspace{-5.0mu}\epsilon_2\mspace{-5.0mu}>$ increasingly negative after successive cycles. In Fig.~\ref{fig:second-run}(d), obtained after run \#2 ended and the sample cooled to room temperature, we see a  well-defined peak in $<\mspace{-5.0mu}\epsilon_2\mspace{-5.0mu}>$ near 3.3 eV.  AFM measurements of the resulting ZnO, shown in Fig. \ref{fig:afm}(a), reveal a RMS roughness of 3.93 nm over a 2x2 $\mu$m$^2$ area.  For comparison, Fig. \ref{fig:afm}(b) shows AFM measurements of a sample prepared without cycling the DEZ. The RMS roughness here is 12.36 nm over a 2x2 $\mu$m$^2$ area.  Ex-situ ellipsometric measurements given in Fig. 5 show that the room-temperature $<\mspace{-5.0mu}\epsilon\mspace{-5.0mu}>$ spectra can be modeled as a bulk ZnO layer 54 nm thick.  Thus a dense layer of ZnO was achieved.  A movie of the realtime spectra from the two runs is available online at \href{http://www4.ncsu.edu/~ejadles}{http://www4.ncsu.edu/~ejadles}.

\begin{figure}
\includegraphics{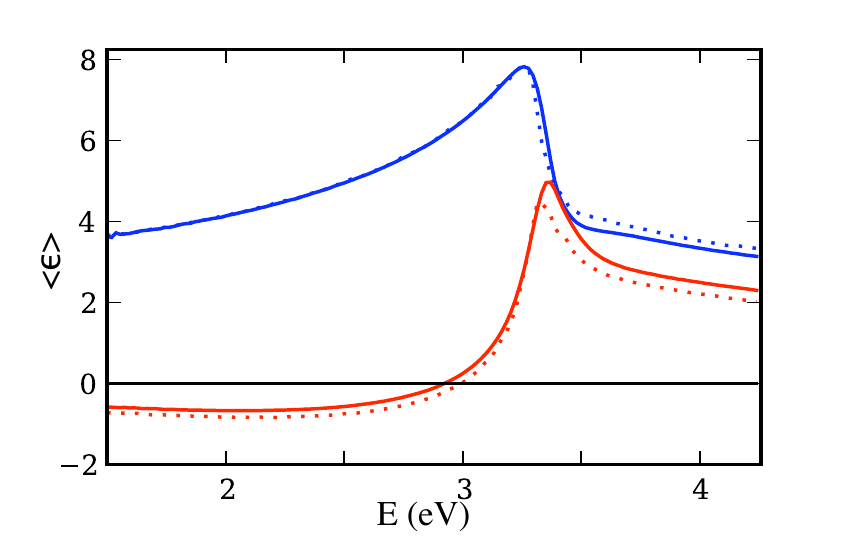}
\caption{Lines: Ex-situ spectroscopic ellipsometric data for the ZnO sample from run \#2. Dotted curves: simulated $<\mspace{-5.0mu}\epsilon\mspace{-5.0mu}>$ data using the dielectric response of bulk ZnO for $\epsilon_o$. (color online) 
\label{fig:se}}
\end{figure}


We now consider the detailed chemistry.  Under ideal conditions, the autocatalytic oxidation reaction of DEZ and O$_2$ is \cite{Ghandhi1980aa}
\begin{equation}
	\mathrm{(C_2 H_5)_2 Zn + 7O_2= ZnO + 4 CO_2 + 5 H_2O}.
	\label{eq:zno}
\end{equation}
However, in practice not all DEZ is oxidized in the gas phase due to kinetic interactions with other gas-phase material.  When unreacted DEZ reaches the substrate surface it thermally decomposes, yielding gaseous Zn, butane, ethane, ethene, and atomic hydrogen. \cite{Dumont1993aa}  The atomic hydrogen results from the thermal decomposition of the ethyl radical. \cite{Heuts1996aa,Gilbert1999aa}  Under these conditions, our data show ZnO removal.  Consistent with this, Kim et al. \cite{Kim2005aa} demonstrate that non-epitaxial ZnO preferentially evaporates, yielding dense two-dimensional material.  However, Kim et al. required temperatures of the order of 800$^\circ$C and processing times of the order of hours, whereas we find reaction times of the order of seconds and temperatures of the order of 450$^\circ$C.  We propose that this is due to atomic hydrogen reacting with ZnO according to
\begin{equation}
	\mathrm{ZnO(s) + 2 H(g) \rightarrow Zn(g) + H_2O(g)},
	\label{eq:sublimation}
\end{equation}
where the hydrogen is generated by decomposition of DEZ.  

During periods of high and low DEZ concentration both ZnO and atomic hydrogen are generated.  However, during periods of high DEZ concentration, more ZnO is deposited than etched, and an excess of atomic hydrogen is generated.  During periods of low DEZ concentration, less ZnO is formed and the residual gas-phase atomic hydrogen preferentially etches non-epitaxially oriented ZnO as well as particulate ZnO in the gas phase.  We take advantage of this by cycling between periods of high and low DEZ concentration to obtain dense 2D material. As added support of the chemical-etching mechanism, we see no ZnO removal at these temperatures when the DEZ concentration is reduced to zero.   Given the complexity of ZnO deposition by MOCVD, in particular the narrow transition window between deposition and etching, we estimate conservatively that the real-time diagnostics capability saved us at least two years of experimentation.


In conclusion, we have shown that particulate ZnO formed in the gas phase can be used as a source of ZnO for deposition.  We have identified two important aspects of ZnO deposition: the need for a seed layer; and the need to manage two exchange currents, one from particulate ZnO, and the other from deposited ZnO.  Finally, we have shown that thermal decomposition of DEZ can preferentially etch ZnO that is either defective or consists of unwanted crystallographic orientations.

\begin{acknowledgments}
This work was supported by DARPA under contract No. W31P4Q-08-1-0003 through the University of Florida.
\end{acknowledgments}


\end{document}